# In-plane anisotropic magnetoresistance in detwinned $BaFe_{2-x}Ni_xAs_2$ ($x$ = 0, 0.6)

Kelly J. Neubauer[1,*], Mason L. Klemm[1], Shirin Mozaffari[2], Lin Jiao[2], Alexei E. Koshelev[3], Alexander Yaresko[4], Ming Yi[1], Luis Balicas[2,**], and Pengcheng Dai[1,***]

1. Department of Physics and Astronomy, Rice University, Houston, TX 77005 USA
2. National High Magnetic Field Laboratory, Florida State University, Tallahassee, FL 32310 USA
3. Materials Science Division, Argonne National Laboratory, Lemont, IL 60439 USA
4. Max Planck Institute for Solid State Research, Heisenbergstrasse 1, 70569 Stuttgart, Germany

**Abstract:**

Understanding the magnetoresistance (MR) of a magnetic material forms the basis for uncovering the orbital mechanisms and charge-spin interactions in the system. Although the parent state of iron-based high-temperature superconductors, including $BaFe_2As_2$, exhibits unusual electron transport properties resulting from spin and charge correlations, there is still valuable insight to be gained by understanding the in-plane MR effect due to twin domains in the orthorhombic antiferromagnetic (AF) ordered state. Here, we study the in-plane magnetoresistance anisotropy in detwinned $BaFe_2As_2$ and compare the results to the non-magnetic Ni-doped sample. We find that in the antiferromagnetically ordered state, $BaFe_2As_2$ exhibits anisotropic MR that becomes large at low temperatures and high fields. Both transverse and longitudinal MRs are highly anisotropic and dependent on the field and current orientations. These results cannot be fully explained by calculations considering only the anisotropic Fermi surface. Instead, the spin orientation of the ordered moment also affects the MR effect, suggesting the presence of a large charge-spin interaction in $BaFe_2As_2$ that is not present in the Ni-doped material.

Magnetoresistance (MR) is the magnetic field dependence of the electric resistance of a given material. Studying the MR effect in a material is an insightful way to understand the sensitivity of the electron scattering to several tunable factors. In magnetic materials, an anomalous MR results from contributions related to the local magnetism and the electronic band structure, which are additive to the ordinary electrical resistance and the conventional orbital MR [1]. An ordinary magnetoresistance (OMR) effect is observed in most metals, including nonmagnetic ones, arising from the simple cyclic motion of electrons under an external applied magnetic field [2]. MR can be measured using transverse or longitudinal configurations, corresponding to a magnetic field perpendicular to current and current along the field, respectively. OMR can exhibit a remarkable transverse effect that typically gives a positive, quadratic field dependence in low fields that saturates at high fields [2]. Longitudinal MR typically has no significant unsaturating field effect. However, simple models do not replicate the MR observed in more complex materials, where the MR can be highly anisotropic due to an anisotropic Fermi surface (FS) or electron (charge)-spin interactions in a magnetically ordered material [2].

Historically, the angular dependence of the MR has been used to investigate the FS of materials. It has been shown that small, anisotropic regions of the FS can induce large quasiparticle scattering and lead to large, anisotropic MR effects under high magnetic fields. Therefore, to determine the field dependence of the MR curve, sensitivities with respect to the shape of the Fermi surface and anisotropic scattering rates must be considered. Additionally, in magnetically ordered materials, spin can play an important role in electron transport due to charge-spin interactions and lead to further anisotropies, including negative values in MR. In general, the resistance is maximal when the magnetization and the electrical current are parallel and minimal when magnetization and current are perpendicular [3].

The electrical transport properties of the iron-based superconductors exhibit anomalous features but understanding the origin of these effects is challenging to do, both quantitatively and qualitatively [4,5]. The parent state of iron-based high-temperature superconductors such as $BaFe_2As_2$ exhibits a tetragonal-to-orthorhombic structural transition at $T_s$ followed by a collinear antiferromagnetic (AFM) order with moment along the *a*-axis of the orthorhombic lattice below the Néel temperature $T_N \approx 140\ (\lesssim T_s)$ K (Figs. 1a and 1b) [6]. Because the orthorhombic state emerges at $T_s$, which is below room temperature, single crystals of $BaFe_2As_2$ form twin domains below $T_s$. To obtain the intrinsic transport properties of $BaFe_2As_2$ within the *ab*-plane, the samples need to be detwinned via the application of a uniaxial strain along one of the in-plane Fe-Fe bond directions [7-10]. Since our recent transport measurements on underdoped superconducting $BaFe_{2-x}Ni_xAs_2$ with AFM order reveal that the upper critical field $H_{c2}$ along the *a*-axis is considerably lower than that along the *b*-axis [11], a determination of the charge-spin interaction induced MR effect in $BaFe_2As_2$ without superconductivity will provide information to unveil the intrinsic in-plane upper critical field anisotropy due solely to superconductivity. In addition, there is great interest in materials with correlations between electronic transport, magnetism, and orbital states whose effects can be effectively tuned for applications in spintronics. Antiferromagnets are very promising spintronics materials given that they display novel effects including anisotropic magnetoresistance (AMR) and tunneling AMR [12-14]. Therefore, a determination of the AMR effect in $BaFe_2As_2$ will shed additional light to our understanding of superconductivity in $BaFe_{2-x}Ni_xAs_2$, and provide information on charge-spin interaction in a collinear AFM ordered magnet.

In this paper, we study the MR effect in the parent $BaFe_2As_2$ and electron doped $BaFe_{2-x}Ni_xAs_2$ materials, which belongs to the iron-based superconductor family (Fig. 1(a)) [6]. In previous studies of the MR effect in the AFM ordered $BaFe_2As_2$, only *c*-axis applied magnetic field MR effect was determined [15-17]. The purpose of our work is to unveil how the AFM order affects the MR effect and determine the electron (charge) spin interactions. To accomplish this goal, we choose single crystals of $BaFe_2As_2$ and $BaFe_{2-x}Ni_xAs_2$ with $x$ = 0.6, where $BaFe_2As_2$ has an orthorhombic AFM ordered ground state with moments along the *a*-axis (Fig. 1(b)) while overdoped $BaFe_{1.4}Ni_{0.6}As_2$ is a non-superconducting paramagnet with a tetragonal crystal structure (Fig. 1(a)). By applying an in-plane magnetic field (Fig. 1(c)) and carrying out transverse and longitudinal MR measurements (Fig. 1(d)) in detwinned $BaFe_2As_2$, and comparing the outcome with similar measurements in nonmagnetic $BaFe_{1.4}Ni_{0.6}As_2$, we can observe conventional orbital mechanism contributions and possible charge-spin interaction induced effects in the MR. We find that in the antiferromagnetically ordered state, $BaFe_2As_2$ exhibits AMR that becomes large at low temperatures and high fields. Both the transverse and longitudinal MRs are highly anisotropic and dependent on the field, current, and magnetization orientations. Similar measurements on nonmagnetic $BaFe_{1.4}Ni_{0.6}As_2$ show no such effect. These results cannot be fully explained by calculations considering only the anisotropy of the Fermi surface. Instead, we conclude that the spin orientation of the ordered moment may also affect the MR effect, particularly in the longitudinal MR, suggesting the presence of a charge-spin interaction in $BaFe_2As_2$.

Single crystals of $BaFe_{2-x}Ni_xAs_2$ ($x$ = 0, 0.6) were grown via the flux method detailed in [18]. Our single crystals formed flat platelets with the *ab*-plane being the basal plane, which we cut into squares with approximately 2-3 mm in length and 0.5 mm in thickness. The quality of the crystals is supported by a residual resistance ratio ($RRR$) of ~ 5 for the resistance measurements ($RRR = \rho_{300\,K}/\rho_{2\,K}$). A $RRR < 10$ is typical for $BaFe_2As_2$ [19]. The Montgomery four-probe method was used to measure the MR with in-plane electrical currents [20,21]. For *c*-axis electrical currents, we used a Corbino geometry to measure the MR [22]. This geometry should allow for the electrical contacts to maximize the cross-sectional area across the *ab*-plane and maintain a consistent electrical density of current flowing along the *c*-axis [11]. A rotator was used to apply magnetic field either along the *a*- or *b*-axis with current either perpendicular or parallel (transverse or longitudinal, respectively) to the field direction. These four configurations with electrical currents in-plane are shown in Fig. 1(d). Further details on the experimental setups used are shown in Fig. A1. Measurements up to 32 T were performed at the NHMFL DC Field Facility, at Florida State University in Tallahassee, FL. All other transport measurements were performed in a Quantum Design 9 T Physical Property Measurement System.

We first consider $BaFe_2As_2$, which has a structural transition from tetragonal to orthorhombic at $T_s$ and magnetic transition below $T_N$ with $T_s \approx T_N \approx 140$ K (Fig. 1(a)) [6]. Upon cooling through the critical temperature $T_s$, structural twins form that make the *ab*-axes indistinguishable [7]. Therefore, we use uniaxial strain to detwin the sample enabling the measurement of the intrinsic in-plane electronic anisotropy. In all cases, a small pressure clamp was used to detwin the sample, shown in Fig. A1(a-b). Below $T_s$ in detwinned samples, the magnetic moments are antiferromagnetically aligned along the *a*-axis as shown in Fig. 1(b). Therefore, when a large magnetic field is applied along the *b*-axis a small spin canting is expected as shown in Fig. 1(c). Since inelastic neutron scattering has revealed a large spin gap of ~10 meV in the AFM ordered state [23-25], a magnetic field of 30 T along the *a*-axis will not be able to induce a spin-flop transition below $T_N$ [26]. Therefore, our in-plane field experiment provides a unique configuration where contributions to the anisotropic MR due to the spin orientation, spin canting, and band structure anisotropy can be observed. Additionally, we compare the $BaFe_2As_2$ samples with the Ni overdoped ones. $BaFe_{2-x}Ni_xAs_2$ displays a superconducting ground state which persists from *x*~0.05-0.25. Above *x* = 0.1, neither the long-range AFM order nor the orthorhombic structure exist [18].

Figures 2(a) and 2(b) shows the temperature dependence of the resistivity of $BaFe_2As_2$ with current flowing along the *a*- and *b*-axes, respectively. In each current configuration, measurements were taken under zero-field and then under an 8.5 T field applied along the *a*- and *b*-axis. Comparison between the zero-field measurements show that current along the *b*-direction leads to a greater effect than current along the *a*-direction. Additionally, resistivity deviates from the high-temperature linear behavior, increasing sharply starting above $T_s$ when current is along the *b*-axis while resistivity decreases over the entire temperature range when current is along the *a*-axis. This is consistent with previous reports [7,27]. The small kink observed at low temperatures is attributed to experimental factors. Additional anisotropy is observed at low temperatures when field is applied transversely to the current. The result is an increase in resistivity below ~ 75 K when $H||a$, $I||b$ and $H||b$, $I||a$. A small decrease in resistivity is observed when current and field are both along the *b* axis while no change is observed when current and field are both along the *a* axis. These results are also shown in the in-plane anisotropy which can be characterized by the ratio of $\rho_b/\rho_a$, as shown in Fig. 2(c). The anisotropy sharply increases above $T_s$ and varies at low temperature dependent on the magnetic field strength and direction. In contrast, no in-plane anisotropy is observed in $BaFe_{2-x}Ni_xAs_2$ (*x* = 0.6) at any temperature as shown in Fig. 2(d-f). These differences are further elucidated in field-dependent MR measurements.

We define the MR effect to be MR = $\Delta\rho/\rho_0 = (\rho(\mu_0 H) - \rho_0)/\rho_0$, where $\rho(\mu_0 H)$ and $\rho_0$ are the resistivities collected under an applied magnetic field of magnitude $\mu_0 H$ and zero field, respectively. Measurements of the MR effect for $BaFe_2As_2$ up to 32 T are shown in Fig. 3. We found that the MR for currents flowing along in-plane directions, in a transverse configuration, was large and positive as shown in Fig. 3(a-b), with the largest effect observed for $\mu_0 H||a$. In both cases, at high fields the field dependence becomes linear and does not saturate. The largest MR response is observed at low temperature.

Upon increasing temperature, the effect decreases until it is fully suppressed at $T_s$. Above $T_s$, the MR is smaller than 0.05% in both configurations.

To analyze our results, we model the magneto-transport of $BaFe_2As_2$ using detailed information about the electronic band structure through Density Functional Theory (DFT) calculations. The band structure we used in these calculations have been experimentally verified through angle resolved photoemission spectroscopy (ARPES) [28-30] and Shubnikov–de Haas Oscillation measurements on detwinned single crystals [31-33]. Our model considers the role of in-plane fields on the magnetoresistivity which to our knowledge have not previously been reported. Figure 3(c) shows the calculated MR effect and the relative anisotropies between both transverse configurations. Our observations indicate a larger magnitude for the MR observed when $\mu_0 H$ ||*a*-axis and *I*||*b*-axis (Fig. 3(a)) relative to the configuration $\mu_0 H$||*b*-axis, *I*||*a*-axis (Fig. 3(b)). Consistent with our calculations, the $\mu_0 H$-linear magnetoresistivity observed at high fields can be reconciled with an orbital response that is inherent to a highly anisotropic Fermi surface [15].

The difference in the magnitude of the MR effect between both transverse configurations is smaller than expected based on our calculations as shown in Fig. 3(a-c). We attribute this difference to additional anisotropy not considered in our calculations due to the spin orientation. Fields along the *a*-axis and currents along the *b*-axis correspond to a configuration having the spin orientation perpendicular to the current direction and this is expected to lead to a minimal contribution to the MR [34]. In contrast, for fields along the *b*-axis and current along the *a*-axis the magnetic moments are oriented nearly along or anti-parallel to the current. This orientation leads to an increased MR implying a reduced anisotropy between both orientations relative to our theoretical calculations.

The measured MRs using longitudinal configurations are shown in Fig. 3(d-e). For current flowing along both axes, the longitudinal MRs are small in comparison to the transverse configurations supporting the main orbital contribution to the transverse magnetoresistance. Despite this, we also observe anisotropic behavior between the two longitudinal configurations. When $\mu_0 H$||*I*||*a*-axis (Fig. 3(d)), there is no clear, systematic MR effect. This is likely due to experimental issues with this specific channel leading to excess noise. When $\mu_0 H$||*I*||*b*-axis (Fig. 3(e)), the MR effect is clearly negative, and the effect decreases with increasing temperature, consistent with a previous study [16,34]. The temperature dependence is highlighted in Fig. A3. Negative longitudinal MR is unusual in a metal. However, a small negative effect has been previously observed in the twinned, doped iron-pnictide samples [16,35], although we did not measure a negative effect in twinned, undoped $BaFe_2As_2$ [Fig. A3(c)]. In our detwinned samples, we observed a larger negative MR when both the current and field are along the *b*-axis. Our calculations, which only consider orbital band effects, do not give negative longitudinal MR (Fig. 3(i)). Since it is known that negative MR can be attributed to the field-induced suppression of magnetic scattering when applying an external field, we must also consider the effect of the spin orientation to correctly describe the longitudinal magnetoresistivity results. Applying an external field along the *b*-axis causes the spins cant away from the *b*-axis. This results in a net magnetization or a greater spin polarization along the *b*-axis direction and the suppression of spin scattering. Negative magnetoresistance is characteristic of ferromagnetic materials and relates to the suppression of electron (charge) spin scattering [36,37]. This small negative slope is apparent in the longitudinal configuration given that it is not subjected to the orbital contribution to the MR. Therefore, we attribute the enhanced negative MR to changes in the spin order which suppresses spin scattering and contributes to the MR. This negative longitudinal MR effect has also been observed in $BaFe_2(As_{1-x}P_x)_2$ where it was attributed to the suppression of spin-fluctuation scattering by a magnetic field [16,35]. Additionally, AFM coupling has been shown to result in a higher resistivity when compared to a ferromagnetic or canted state in other materials including FeRh [38].

Additionally, we measured the magnetoresistivity for in-plane magnetic fields and currents along the *c*-axis, as shown in Figs. 3(g-h) under fields up to 9 T. In this orientation, the current and spin orientation are always perpendicular to each other making it distinct from the transverse orientations shown in Figs. 3(a-f), where field, spin orientation, and currents are all oriented within the plane. The MR effect is large with a greater magnitude when $\mu_0 H || a$-axis (Fig. 3(g)) when compared to $\mu_0 H || b$-axis (Fig. 3(h)). As with in-plane currents, the effect is reduced with increasing temperature. Our experimental results are qualitatively consistent with the calculations as shown in Fig. 3(i) that expects anisotropy in the MR resulting from the anisotropy of the Fermi surface. Compared to our calculations, the anisotropy is even greater (Fig. A4). This can be reconciled considering the effect of spin canting when the field is applied along the *b*-axis in contrast to the lack of spin canting when the field is applied along the *a*-axis. A smaller canting corresponds to a more anti-parallel or antiferromagnetic alignment of the spins which leads to a higher resistance. When the spins are slightly canted, the resistance decreases therefore increasing the anisotropy between both field orientations. This effect has been observed in other materials such as $Sr_2IrO_4$ where the small canting of the spins due to planar fields results in a large anisotropy in the magnetoresistance associated to inter-planar currents [39].

We measured the angular dependent AMR which is defined as MR $= (\rho(\mu_0 H(\theta)) - \rho(\mu_0 H(\theta = 0)))/\rho(\mu_0 H(\theta = 0))]$ where $\theta$ is the angle between the magnetic field and the *a*-axis (see Fig. A1(d)). The temperature dependence of the AMR is shown in Fig. 4(a-b) with current along the *b*- (Fig. 4(a)) and *a*-axis (Fig. 4(b)) with a magnetic field of 8.5 T. The maximal effect occurs when the field is perpendicular to the current, or at $\theta = 90°$. The small asymmetry observed can be attributed to experimental effects of slight misalignment of the field. The effect is largest at low temperatures and fully suppressed when the temperatures are raised up to $T_s$. The field dependence is shown in Figs. 4(c-d) with current applied along the *b*- (Fig. 4(c)) and *a*-axis (Fig. 4(d)) at a temperature of 10 K. The effect increases with increasing magnetic field. The anisotropic MR effect is below the resolution limit for all angles in $BaFe_{2-x}Ni_xAs_2$ ($x$ = 0.6) as shown in Fig. 4(e-f). The angular dependence of the AMR follows a two-fold symmetric curve similar to that observed in other ferromagnetic metals and canted AFM materials including $La_{0.4}Sr_{0.6}MnO_3$ [40,41]. It reflects the strength of the Lorentz force which is minimal at $\theta = 0°$ and maximal at $\theta = 90°$, albeit for the $I || b$-axis one observes a pronounced negative AMR as the field increases in contrast to the positive AMR observed for the *I*||*a*-axis.

In conclusion, we reported on the magnetoresistivity of $BaFe_{2-x}Ni_xAs_2$, showing the clear anisotropic effects when an in-plane magnetic field is applied to detwinned $BaFe_2As_2$ which displays a collinear AFM structure. By comparing our experimental results to our calculations, we demonstrated that the anisotropic MR is driven by ordinary orbital mechanisms in addition to charge-spin interactions. Alternatively, the mechanism behind the unique effect is a combination of scattering caused by the anisotropy of the Fermi surface and that due to the spin orientation and its evolution under a magnetic field. The relationship between these mechanisms provides additional insight towards our understanding of the MR effect in magnetic ordered materials and provides the basis to understand the in-plane superconducting gap anisotropy in AFM ordered iron pnictide superconductors.

The experimental and basic materials synthesis work at Rice is supported by the U.S. DOE, BES under Grant No. DE-SC0012311 and by the Robert A. Welch Foundation under Grant No. C-1839, respectively (P.D.). L.B. is supported by the US DoE, Basic Energy Sciences program through award DE-SC0002613. A portion of this work was performed at the National High Magnetic Field Laboratory, which is supported by National Science Foundation Cooperative Agreement No. DMR-1644779 and the State of Florida. Work of A.E.K at Argonne National Laboratory was funded by the US Department of Energy, Office of Science, Basic Energy Sciences, Materials Sciences and Engineering Division.

## Figures

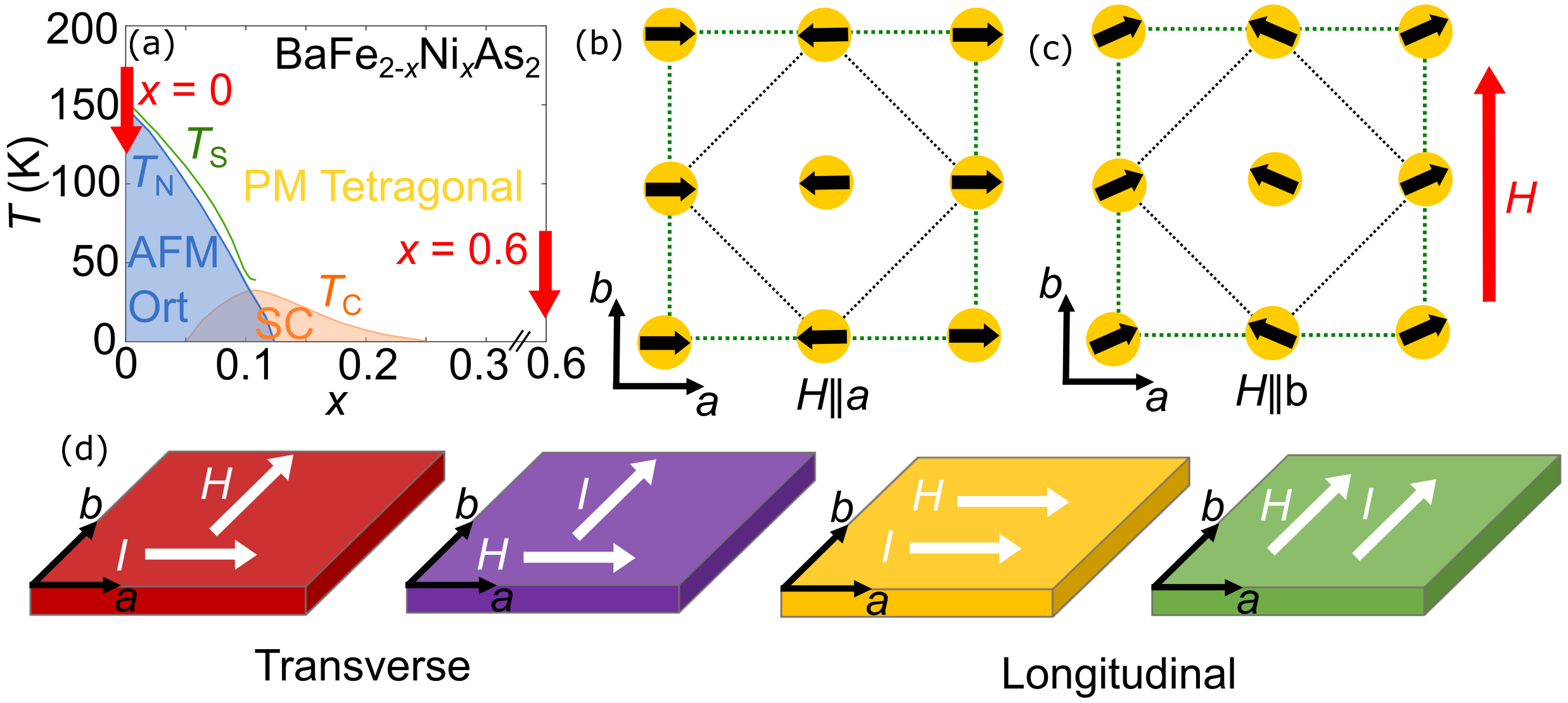

FIG. 1. (a) Phase diagram of $BaFe_{2-x}Ni_xAs_2$ adapted from [6]. (b) Low temperature spin configuration of $Fe^{2-}$ ions on the FeAs plane for a detwinned $BaFe_2As_2$ crystal under zero field and for $\mu_0 H||a$-axis. (c) Spin canted structure with a net magnetization along the *b*-axis when $\mu_0 H||b$-axis. (d) Experimental configurations for MR measurements with in-plane field and current either along the crystallographic *a* or *b* axes.

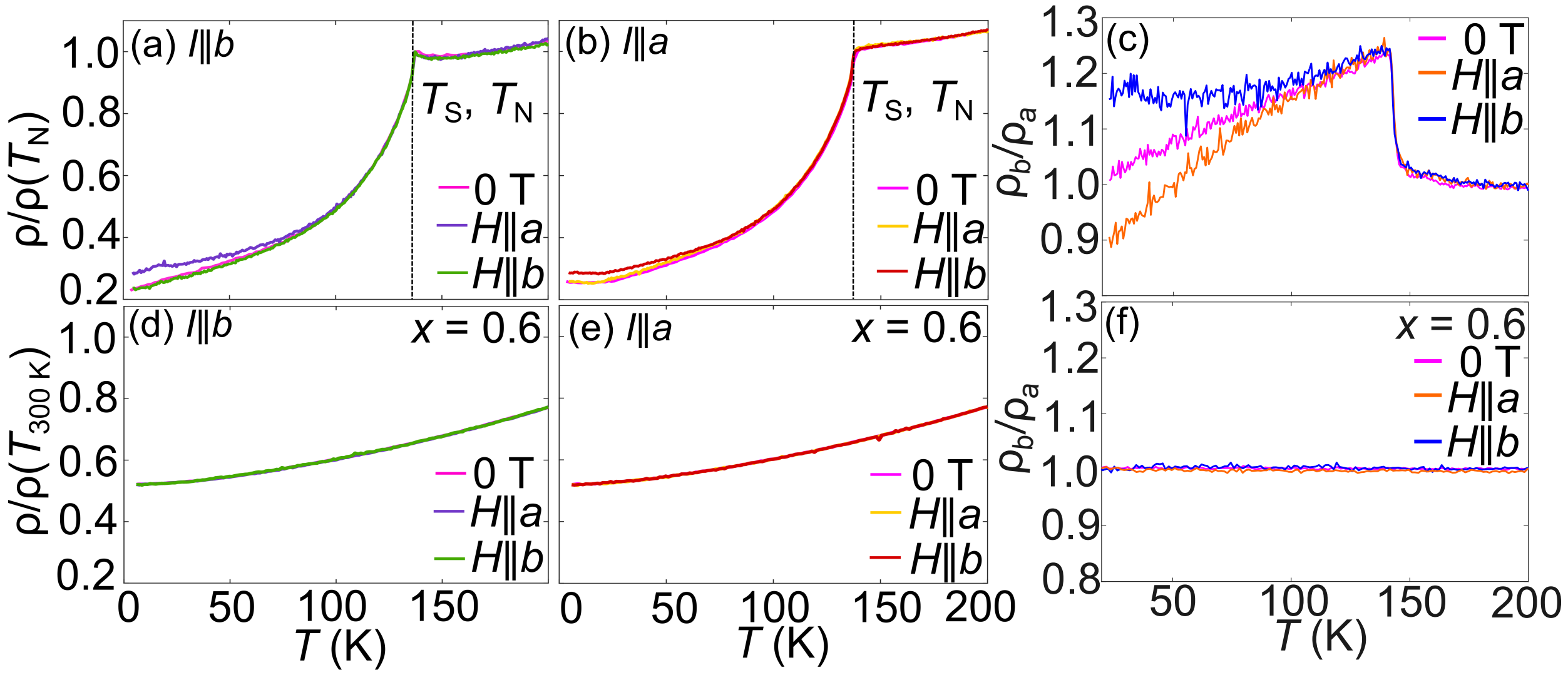

FIG. 2. (a) Temperature dependence of the in-plane resistivity for currents flowing along the *b*-axis direction. (b) Temperature dependence of the in-plane resistivity with currents applied along the *a*-axis direction. (c) Temperature dependence of the anisotropy of the in-plane resistivity. (d) Temperature dependence of the resistivity for an over-doped sample of $BaFe_{2-x}Ni_xAs_2$, with $x$ = 0.6 and current along the *b*-axis. (e) Temperature dependence of the resistivity from an over-doped sample of $BaFe_{2-x}Ni_xAs_2$, with $x$ = 0.6 and current flowing along the *a*-axis. In all plots the magnetic field is 8.5 T. (f) Temperature dependence of the anisotropy of the in-plane resistivity for an over-doped sample of $BaFe_{2-x}Ni_xAs_2$, with $x$ = 0.6.

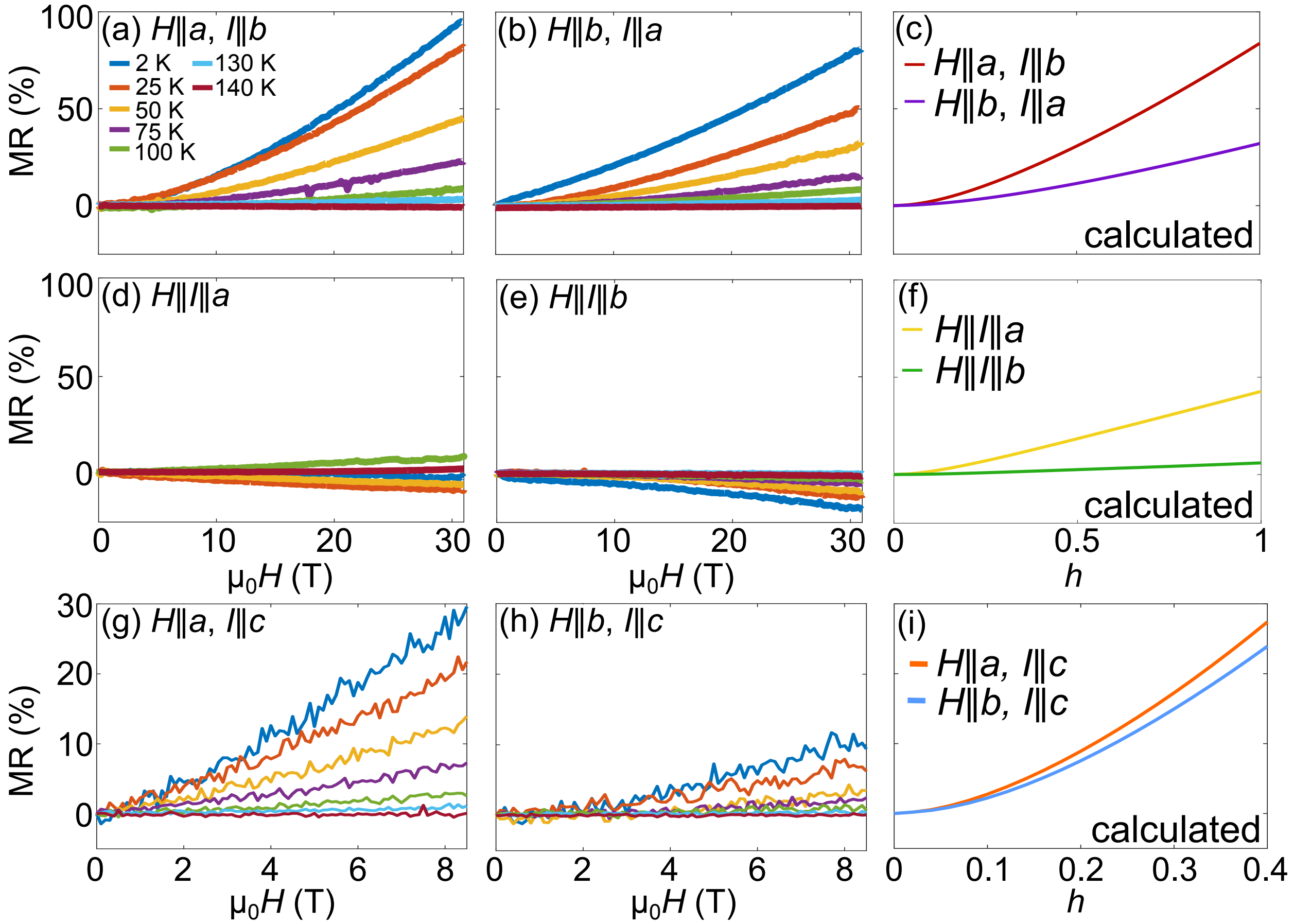

FIG. 3. (a) Magnetoresistance of a detwinned $BaFe_2As_2$ crystal for $\mu_0H||a$-axis and $I||b$-axis. (b) Magnetoresistance of a detwinned $BaFe_2As_2$ crystal for $\mu_0H||b$-axis and $I||a$-axis. (c) Calculated transverse MR for currents flowing along a planar direction. (d) Magnetoresistance of a detwinned $BaFe_2As_2$ crystal for $\mu_0H||a$-axis and $I||a$-axis. (e) Magnetoresistance of a detwinned $BaFe_2As_2$ crystal for $\mu_0H||b$-axis and $I||b$-axis. (f) Calculated longitudinal MR. (g) Magnetoresistance of a detwinned $BaFe_2As_2$ crystal for $\mu_0H||a$-axis and $I||c$-axis. (h) Magnetoresistance of a detwinned $BaFe_2As_2$ crystal for $\mu_0H||b$-axis and $I||c$-axis. (i) Calculated transverse MR for $I||c$-axis.

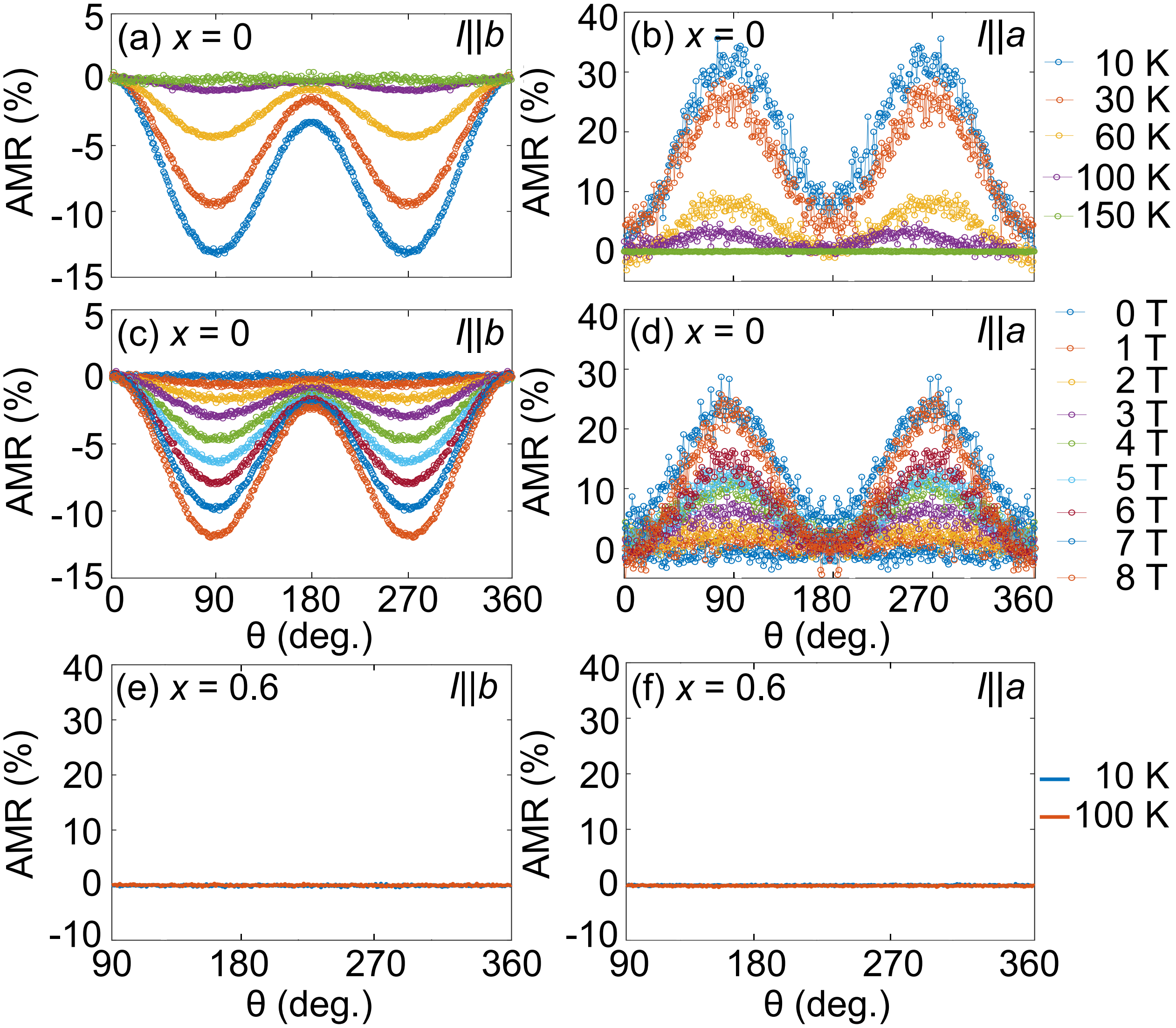

FIG. 4. (a) Temperature dependence of the angular magnetoresistance (AMR) for a detwinned $BaFe_2As_2$ crystal with *I*||*b* under $\mu_0 H$ = 8.5 T. (b) Temperature dependence of the AMR for a detwinned $BaFe_2As_2$ crystal with *I*||*a*-axis for $\mu_0 H$= 8.5 T. (c) Field dependence of the AMR for a detwinned $BaFe_2As_2$ crystal when *I*||*b*-axis at 10 K. (d) Field dependence of the AMR for a detwinned $BaFe_2As_2$ crystal when *I*||*a*-axis at 10 K. (e) Temperature dependence of the AMR in an over-doped $BaFe_{2-x}Ni_xAs_2$ sample having *x*=0.6 when *I*||*b*-axis under $\mu_0 H$ = 8.5 T. (f) Temperature dependence of the AMR for an over-doped $BaFe_{2-x}Ni_xAs_2$ sample having *x* = 0.6 when *I*||*a*-axis under $\mu_0 H$ = 8.5 T.

**Appendix A: Experimental Details and Results**

The experimental setup used to detwin the sample, apply field, and measure the resistivity is shown in Fig. A1.

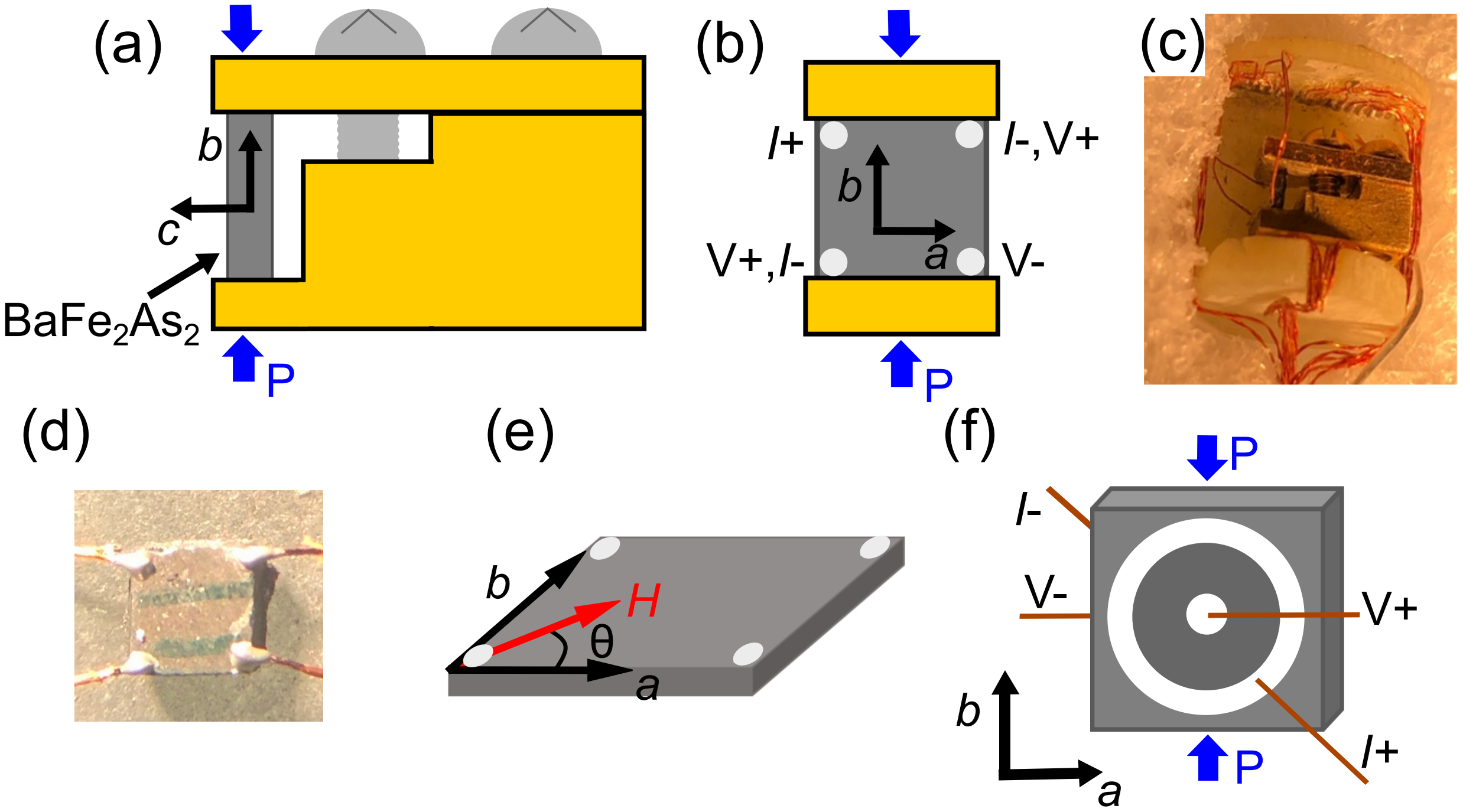

FIG. A1. (a) Schematic of the strain device and sample. (b) Schematic of the strain device and the configuration for the in-plane magnetoresistivity. (c) Photograph of the strain device and sample with in-plane current setup. (d) Zoomed-in photograph of the 2x2 mm sample with in-plane current setup. (e) Schematic of the angular dependence of the AMR measurements, indicating the origin of the angle $\theta$. (f) Schematic of the Corbino configuration for the *c*-axis current magnetoresistivity. The back surface of the sample is identically wired to the front.

To show consistency, we compare our MR results obtained in fields up to 8.5 T (LF) to those obtained up to 31 T (HF) in Fig A2(a-d). The magnitude and field dependence of the results are in good agreement with each other.

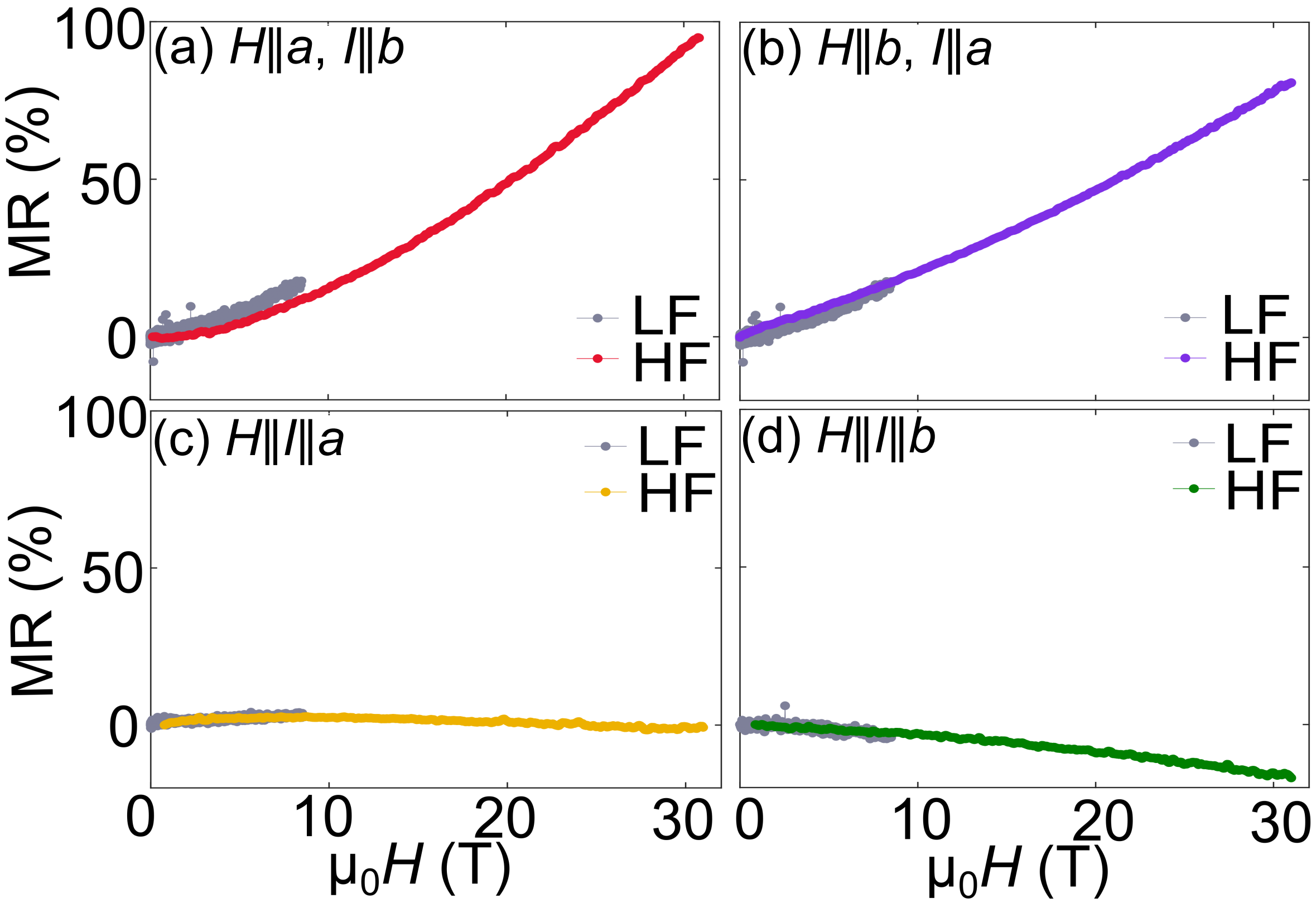

FIG. A2. (a) In-plane MR of a detwinned $BaFe_2As_2$ sample from the 9 T Physical Property Measurement System (PPMS) (gray) compared to those collected in a 32 T Bitter resistive magnet (red) with $\mu_0H||a$-axis and $I||b$-axis. (b) In-plane MR of a detwinned $BaFe_2As_2$ sample from the 9 T PPMS (gray) compared to the 32 T Bitter resistive magnet (purple) with $\mu_0H||b$-axis and $I||a$-axis. (c) In-plane MR for a detwinned $BaFe_2As_2$ crystal collected from the 9 T PPMS (gray) compared to 32 T Bitter resistive magnet (yellow) with $\mu_0H||a$-axis and $I||a$-axis. (d) In-plane MR of a detwinned $BaFe_2As_2$ crystal from the 9-T PPMS (gray) compared to the 32 T Bitter resistive magnet (green) with $\mu_0H||b$-axis and $I||b$-axis.

In Fig. A3, we show the longitudinal MR results from Fig. 3(d-e) with an enlarged scale to show the observed temperature dependence clearly. The MR measured for $\mu_0H||I||a$ provides a less reliable temperature dependence. We believe this results from experimental connectivity issues observed in this particular channel. Additionally, we include several measurements on a twinned sample with $\mu_0H||b$-axis and $I||b$-axis. We observe no negative longitudinal MR in the twinned sample in contrast to the detwinned sample.

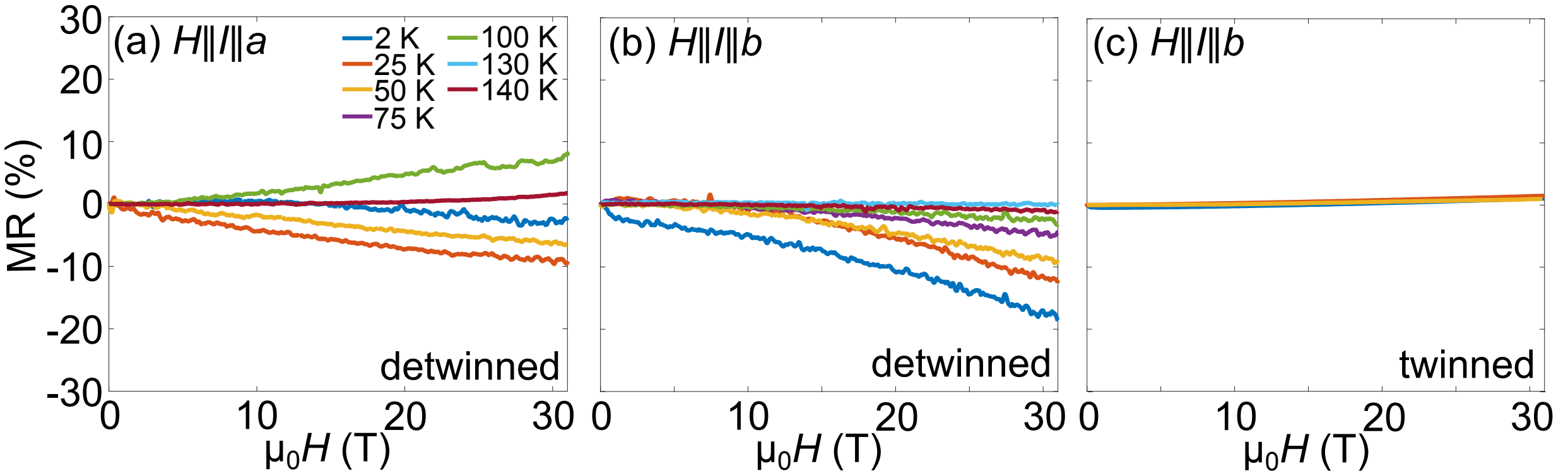

FIG. A3. (a) In-plane longitudinal MR of a detwinned $BaFe_2As_2$ sample collected in a 32 T Bitter resistive magnet with $\mu_0 H||a$-axis and $I||a$-axis. (b) In-plane longitudinal MR of a detwinned $BaFe_2As_2$ sample collected in a 32 T Bitter resistive magnet with $\mu_0 H||b$-axis and $I||b$-axis. (c) In-plane longitudinal MR of a twinned $BaFe_2As_2$ sample collected in a 32 T Bitter resistive magnet with $\mu_0 H||b$-axis and $I||b$-axis.

In Fig. A4, we provide a representative comparison of the experimental data (solid lines) to calculations (dashed line).

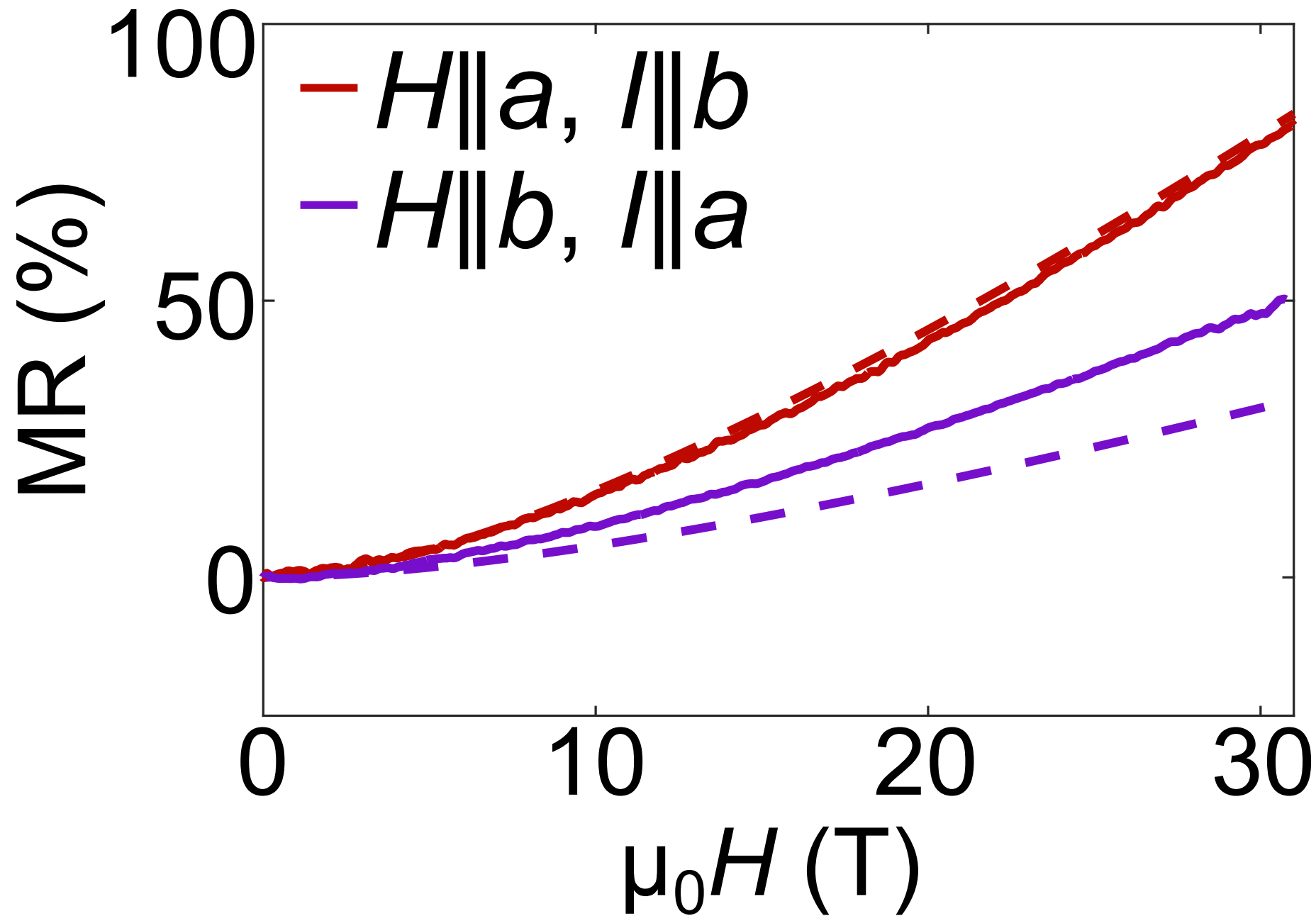

FIG. A4. (a) In-plane transverse MR of a detwinned $BaFe_2As_2$ sample collected in a 32 T Bitter resistive magnet at 25 K. Solid lines represent experimental data. Dashed lines represent calculations.

**Appendix B: Calculation Details**

We estimate the spin canting angle and its field and temperature dependence by using a nearest-neighbor Heisenberg model with single-site anisotropy approximation, as explained in Reference [26]. This allows for an approximation of the spin canting of the Fe atoms away from the *a*-axis when the field is applied along the *b*-axis. From base temperature (2 K) neutron scattering results, the magnetic moment $\mu$ ~1 $\mu_B$ and the nearest neighbor magnetic exchange coupling ~60 meV [24,25]. We used this information and the temperature dependence of the magnetic susceptibility measurements in Reference [42] to estimate the spin canting angle of ~0.3 degrees at 12 T. We estimate the field dependence of the spin canting at 2 K in Fig. B1. Overall, these results indicate that the spin canting angle is small (<1 degree) in our experiments.

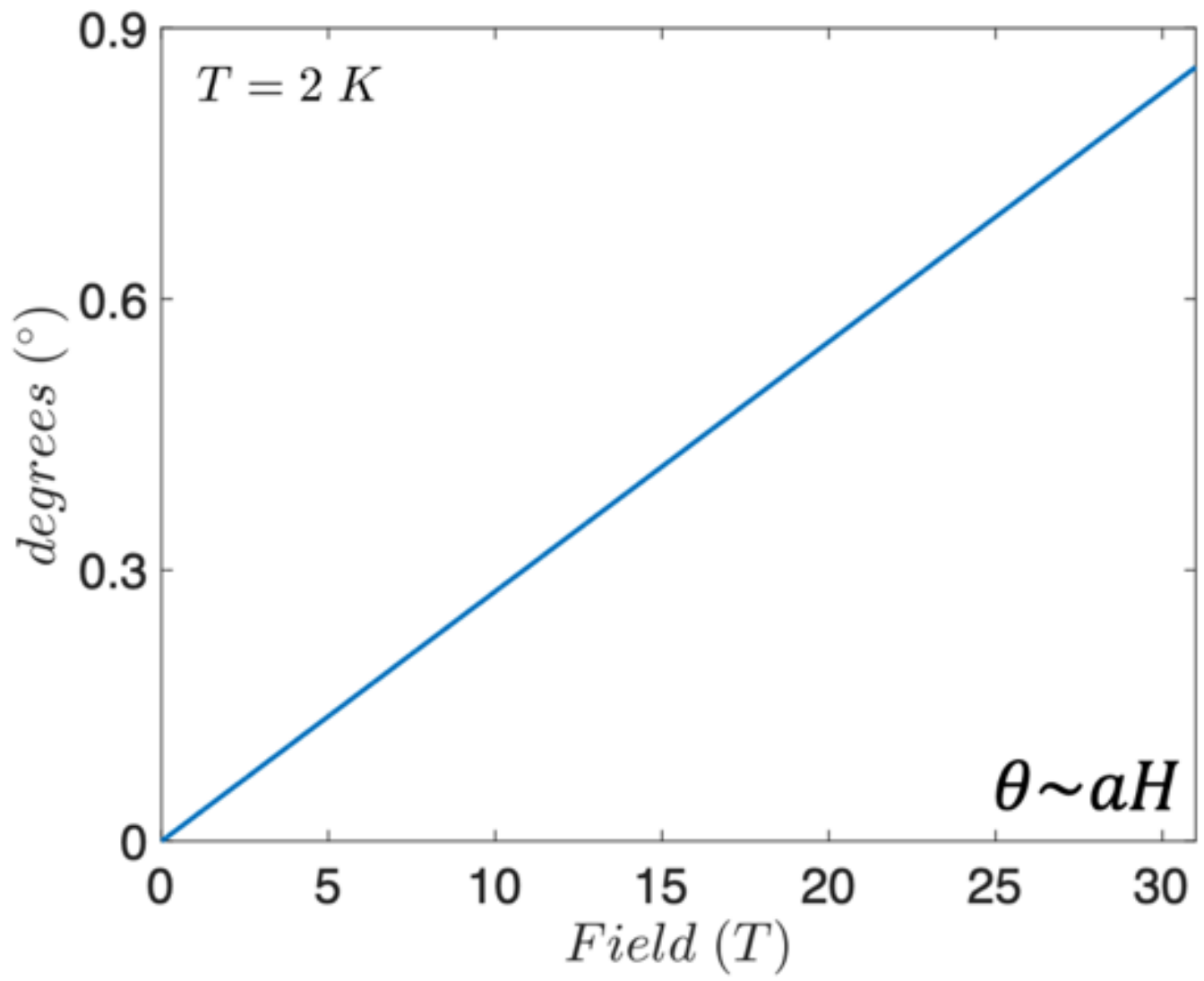

FIG. B1. Estimated field dependence of spin canting calculated using Heisenberg model approximation.

*email: kelly.j.neubauer@rice.edu
**email: balicas@magnet.fsu.edu
***email: pdai@rice.edu

## References

[1] T. Kasuya, Prog. Theor. Phys. **16**, 58 (1956).

[2] A. B. Pippard, Magnetoresistance in metals, Cambridge University Press Cambridge, (1989).

[3] M. Getzlaff, Fundamentals of Magnetism Springer-Verlag Berlin Heidelberg (2008).

[4] D. C. Johnston, Adv. Phys. **59**, 803 (2010).

[5] R. M. Fernandes, A. I. Coldea, H. Ding, I. R. Fisher, P. J. Hirschfeld, and G. Kotliar, Nature **601**, 35 (2022).

[6] P. Dai, Rev. Mod. Phys. **87**, 855 (2015).

[7] I. R. Fisher, L. Degiorgi, and Z. X. Shen, Rep. Prog. Phys. **74**, 124506 (2011).

[8] J.-H. Chu, J. G. Analytis, K. De Greve, P. L. McMahon, Z. Islam, Y. Yamamoto, and I. R. Fisher, Science **329**, 824 (2010).

[9] M. A. Tanatar, N. Ni, S. L. Bud'ko, P. C. Canfield, and R. Prozorov, Supercond. Sci. Technol. **23**, 054002 (2010).

[10] H. Man et al., Phys. Rev. B **92**, 134521 (2015).

[11] M. Klemm et al., arXiv:2311.07697 (2023).

[12] I. Fina et al., Nat. Commun. **5**, 4671 (2014).
[13] B. G. Park et al., Nat. Mater. **10**, 347 (2011).
[14] X. Martí et al., Phys. Rev. Lett. **108**, 017201 (2012).
[15] N. Maksimovic, I. M. Hayes, V. Nagarajan, J. G. Analytis, A. E. Koshelev, J. Singleton, Y. Lee, and T. Schenkel, Phys. Rev. X **10**, 041062 (2020).
[16] I. M. Hayes, Z. Hao, N. Maksimovic, S. K. Lewin, M. K. Chan, R. D. McDonald, B. J. Ramshaw, J. E. Moore, and J. G. Analytis, Phys. Rev. Lett. **121**, 197002 (2018).
[17] L. Jiao, Z. F. Weng, X. Y. Tang, L. K. Guo, T. Shang, L. Yang, H. Q. Yuan, Y. Y. Wu, and Z. C. Xia, J. Korean Phys. Soc. **63**, 453 (2013).
[18] Y. Chen, X. Lu, M. Wang, H. Luo, and S. Li, Supercond. Sci. Technol. **24**, 065004 (2011).
[19] A. S. Sefat et al., Phys. Rev. B **79**, 094508 (2009).
[20] H. C. Montgomery, J. Appl. Phys. **42**, 2971-2975, (1971).
[21] C. A. M. dos Santos, A. de Camps, M. S. da Luz, B. D. White, J. J. Neumeier, B. S. de Lima, and Y. Shigue, J. Appl. Phys. **110**, 083703, (2011).
[22] X. F. Wang, T. Wu, G. Wu, H. Chen, Y. L. Xie, J. J. Ying, Y. J. Yan, R. H. Liu, and X. H. Chen, Phys. Rev. Lett. **102**, 117005 (2009).
[23] K. Matan, R. Morinaga, K. Iida, and T. J. Sato, Phys. Rev. B **79**, 054526 (2009).
[24] L. W. Harriger, H. Q. Luo, M. S. Liu, C. Frost, J. P. Hu, M. R. Norman, and P. Dai, Phys. Rev. B **84**, 054544 (2011).
[25] C. Wang, R. Zhang, F. Wang, H. Luo, L. P. Regnault, P. Dai, and Y. Li, Phys. Rev. X **3**, 041036 (2013).
[26] J. Maiwald, I. I. Mazin, and P. Gegenwart, Phys. Rev. X **8**, 011011 (2018).
[27] J.-H. Chu, J. G. Analytis, D. Press, K. De Greve, T. D. Ladd, Y. Yamamoto, and I. R. Fisher, Phys. Rev. B **81**, 214502 (2010).
[28] M. D. Watson et al., npj Quantum Mater. **4**, 36 (2019).
[29] M. Fuglsang Jensen, V. Brouet, E. Papalazarou, A. Nicolaou, A. Taleb-Ibrahimi, P. Le Fèvre, F. Bertran, A. Forget, and D. Colson, Phys. Rev. B **84**, 014509 (2011).
[30] H. Pfau, C. R. Rotundu, J. C. Palmstrom, S. D. Chen, M. Hashimoto, D. Lu, A. F. Kemper, I. R. Fisher, and Z. X. Shen, Phys. Rev. B **99**, 035118 (2019).
[31] T. Terashima et al., Phys. Rev. Lett. **107**, 176402 (2011).
[32] J. G. Analytis, J. H. Chu, R. D. McDonald, S. C. Riggs, and I. R. Fisher, Phys. Rev. Lett. **105**, 207004 (2010).
[33] P. F. S. Rosa et al., Phys. Rev. B **90**, 195146 (2014).
[34] F. Duan and J. Guojun, Introduction to Condensed Matter Physics, World Scientific (2005).
[35] F. Rullier-Albenque, D. Colson, and A. Forget, Phys. Rev. B **88**, 045105 (2013).
[36] H. Yamada and S. Takada, Prog. Theor. Phys. **48**, 1828 (1972).
[37] J. Smit, Physica **17**, 612 (1951).
[38] V. L. Moruzzi and P. M. Marcus, Phys. Rev. B **46**, 14198 (1992).
[39] C. Wang, H. Seinige, G. Cao, J. S. Zhou, J. B. Goodenough, and M. Tsoi, Phys. Rev. X **4**, 041034 (2014).
[40] M. Egilmez, M. M. Saber, A. I. Mansour, R. Ma, K. H. Chow, and J. Jung, Appl. Phys. Lett. **93**, 182505 (2008).
[41] J.-B. Yau et al., J. Appl. Phys. **102**, 103901 (2007).

[42] M. He, L. Wang, F. Ahn, F. Hardy, T. Wolf, P. Adelmann, J. Schmalian, I. Eremin, and C. Meingast, Nat. Commun **8**, 504 (2017).